\documentclass{jpp}

\usepackage{graphicx}
\usepackage{natbib}

\ifCUPmtlplainloaded \else
  \checkfont{eurm10}
  \iffontfound
    \IfFileExists{upmath.sty}
      {\typeout{^^JFound AMS Euler Roman fonts on the system,
                   using the 'upmath' package.^^J}%
       \usepackage{upmath}}
      {\typeout{^^JFound AMS Euler Roman fonts on the system, but you
                   dont seem to have the}%
       \typeout{'upmath' package installed. JPP.cls can take advantage
                 of these fonts, if you use 'upmath' package.^^J}%
      }
  \else
  \fi
\fi

\ifCUPmtlplainloaded \else
  \checkfont{msam10}
  \iffontfound
    \IfFileExists{amssymb.sty}
      {\typeout{^^JFound AMS Symbol fonts on the system, using the
                'amssymb' package.^^J}%
       \usepackage{amssymb}%
         \let\leq=\leqslant
         \let\geq=\geqslant
      }{}
  \fi
\fi

\ifCUPmtlplainloaded \else
  \IfFileExists{amsbsy.sty}
    {\typeout{^^JFound the 'amsbsy' package on the system, using it.^^J}%
     \usepackage{amsbsy}}
    {}
\fi

\newsavebox{\astrutbox}
\sbox{\astrutbox}{\rule[-5pt]{0pt}{20pt}}

\title[Nature and dynamics of wave overreflection]{Nature and dynamics of overreflection of Alfv${\rm \acute{e}}$n waves in
MHD shear flows}

\author[D. Gogichaishvili, G. Chagelishvili, R. Chanishvili and J. Lominadze]%
{D.\ns G\ls O\ls G\ls I\ls C\ls H\ls A\ls I\ls S\ls H\ls V\ls I\ls
L\ls I$^{1}$,\ns G.\ns C\ls H\ls A\ls G\ls E\ls L\ls I\ls S\ls H\ls
V\ls I\ls L\ls I$^{2,3}$,\ns \break R.\ns C\ls H\ls A\ls N\ls I\ls
S\ls H\ls V\ls I\ls L\ls I$^{2,3}$
 \and J.\ns L\ls O\ls M\ls I\ls N\ls A\ls D\ls Z\ls E$^{2,3}$\thanks{Deceased}}

\affiliation{$^1$Department of
Physics, The University of Texas, Austin, TX 78712, USA\\[\affilskip]
$^2$Abastumani Astrophysical Observatory, Ilia State University,
Tbilisi, 0162, Georgia\\[\affilskip]
$^3$Institute of Geophysics, Tbilisi State University, Tbilisi,
0193, Georgia}

\pubyear{2013} \volume{0} \pagerange{0--0}
\date{?; revised ?; accepted ?. - To be entered by editorial office}
\begin{document}

\maketitle

\begin{abstract}
Our goal is to gain new insights into the physics of wave
overreflection phenomenon in MHD nonuniform/shear flows changing the
existing trend/approach of the phenomenon study. The performed
analysis allows to separate from each other different physical
processes, grasp their interplay and, by this way, construct the
basic physics of the overreflection in incompressible MHD flows with
linear shear of mean velocity, ${\bf U}_0=(Sy,0,0)$, that contain
two different types of Alfv${\rm \acute{e}}$n waves. These waves are
reduced to pseudo- and shear shear-Alfv${\rm \acute{e}}$n waves when
wavenumber along $Z$-axis equals zero (i.e., when $k_z=0$).
Therefore, for simplicity, we labelled these waves as: P-Alfv${\rm
\acute{e}}$n and S-Alfv${\rm \acute{e}}$n waves (P-AWs and S-AWs).
We show that: (1) the linear coupling of counter-propagating waves
determines the overreflection, (2) counter-propagating P-AWs are
coupled with each other, while counter-propagating S-AWs are not
coupled with each other, but are asymmetrically coupled with P-AWs;
S-AWs do not participate in the linear dynamics of P-AWs, (3) the
transient growth of S-AWs is somewhat smaller compared with that of
P-AWs, (4) the linear transient processes are highly anisotropic in
wave number space, (5) the waves with small streamwise wavenumbers
exhibit stronger transient growth and become more balanced, (6)
maximal transient growth (and overreflection) of the wave energy
occurs in the two-dimensional case -- at zero spanwise wavenumber.

To the end, we analyze nonlinear consequences of the described
anisotropic linear dynamics -- they should lead to an anisotropy of
nonlinear cascade processes significantly changing their essence,
pointing to a need of revisiting the existing concepts of cascade
processes in MHD shear flows.
\end{abstract}


\section{Introduction}
Nonuniform flows are ubiquitous both in nature and in laboratory.
They occur in atmospheres, oceans, solar wind, stars, astrophysical
disks, pipe flows, tokamak reactors, etc. Complex dynamics of these
systems is, in many respects, a consequence of their nonuniform
kinematics. One of the basic manifestations of flow shear is wave
overreflection phenomenon -- substantial growth of
counter-propagating wave perturbations -- that occurs universally
whenever flow has non-zero shear. For instance, in astrophysical
discs, there is the overreflection of spiral-density waves, in the
atmosphere -- internal-gravity waves, in the solar wind -- MHD
waves, etc.

The essence of the overreflection is the following. If initially in
a shear flow exist just waves having definitely directed group
velocity and, consequently, propagating in one direction (i.e.
counter-propagating waves are absent initially), in the course of
time, appears and substantially increase counter-propagating waves,
i.e., waves that have oppositely (to the initial waves) directed
group velocity. Of course, if the initial/incident waves are
localized in the flow shear direction, the overreflection phenomenon
has visual manifestation -- it appears reflected (also localized in
the shear direction) waves having larger, then incident waves,
amplitude. However, in the non-modal approach, investigating the
dynamics of spatial Fourier harmonics of waves (so called, Kelvin
waves) that are not localized in the physical space, the visual
manifestation is absent and one can mathematically describe the
overreflection by the appearance and increase of counter-propagating
waves, i.e. -- waves having oppositely directed group velocity. We
aim to go deeper into the physical mechanism responsible for this
overreflection in an incompressible MHD shear flow.

The overreflection phenomenon is usually analysed on the basis of a
single second-order ordinary differential (wave) equation
\citep[e.g., see seminal papers on overreflection][]{Goldreich78,
Goldreich79, Lindzen85, Farrell93a}. This approach describes
counter-propagating waves by a single/physical variable and,
consequently, possible dynamical processes between these waves
(e.g., their coupling) are, in fact, left out of consideration. To
fill this gap -- to reduce the perturbation equations to the set of
first order differential equations for individual
counter-propagating wave -- we propose, at first, to find
``eigen-variable'' for each wave component. I.e., to find variables
for which the linear matrix operator is diagonal in the shearless
limit. Then, one have to generalize the eigen-variables for non-zero
shear case. The generalization is not unique possible procedure. But
the optimal one is quite easily findable. In the considered MHD
flow, the generalized variables are the renormalized
Els$\ddot{a}$sser variables (see equations 2.12). Resulting first
order ordinary differential equations (written for the generalized
variables) separate from each other different physical processes,
make it possible to grasp their interplay and understand the basic
physics of the overreflection phenomenon. In the renormalized
Els$\ddot{a}$sser variables we get two different types of Alfv${\rm
\acute{e}}$n waves in incompressible MHD flows with linear shear of
mean velocity, ${\bf U}_0=(Sy,0,0)$. As it is mentioned in the
abstract, these waves are reduced to pseudo- and shear-Alfv${\rm
\acute{e}}$n waves when wavenumber along $Z$-axis equals zero (i.e.,
when $k_z=0$). Therefore, for simplicity, we labelled these waves
as: P-Alfv${\rm \acute{e}}$n and S-Alfv${\rm \acute{e}}$n waves
(P-AWs and S-AWs). We carried out analytical (Kelvin mode) and
numerical analysis of the linear dynamics of P-AWs and S-AWs. Our
analysis has clearly demonstrated linear coupling between these
waves. We describe in detail transient growth of counter-propagating
components of P-AWs and S-AWs when only one of them exists initially
in the flow. The amplification of these (so-called, transmitted and
reflected) waves, i.e. \emph{the wave overreflection phenomenon}, is
determined by the linear coupling of the waves induced by shear flow
non-normality.

Historically, the non-normality of shear flows considerably delayed
the full understanding of their behavior. In fact, this feature and
its dynamical consequences only became well understood by the
hydrodynamic community in the 1990s \citep[e.g., see][]{Reddy93,
Trefethen93, Schmid07}. Shortcomings of the traditional modal
analysis (i.e., spectral expansion of perturbations in time and then
analysis of eigenfunctions) for shear flows have been revealed and
an alternative Kelvin wave approach -- a special kind of the
non-modal approach -- has become well established and has been
extensively used since the 1990s. Kelvin waves represent the basic
``elements'' of dynamical processes at constant shear rate
\citep{Chagelishvili96,Yoshida05} and greatly help to understand
finite-time transient phenomena in shear flows. In particular, it
reveals a channel of linear coupling among different branches/modes
of perturbations in shear flows \citep{Chagelishvili97a,
Chagelishvili97b}, leading to energy exchange between vortex and
wave modes and between different wave branches.

A new, \emph{bypass transition}, concept was also formulated by the
the hydrodynamic community to explain the onset of turbulence in
spectrally stable shear flows \citep[e.g., see][and references
therein]{Baggett95, Grossmann00, Eckhardt07} on the basis of the
interplay between linear transient growth and nonlinear positive
feedback. The bypass scenario differs fundamentally from the
classical turbulence scenario, which is based on exponentially
growing perturbations in a system that supplies turbulent energy. In
the classical case, the nonlinearity is not vital to the existence
of the perturbations. Instead, it merely determines their scales,
via the direct/inverse cascade.

This breakthrough led to renewed comprehension of different aspects
of shear flow dynamics. This paper, being along these lines, aims to
provide new insight into the physics of wave overreflection on the
example of P-AWs and S-AWs in incompressible MHD flows with linear
shear of mean velocity profile. Actually, our
study can be of wide applicability:\\
(i) the method of characterizing overreflection presented here is
optimal, moreover, canonical, and can be easily applied for deeper
understanding of the overreflection phenomenon in other important
cases, such as spiral-density waves in astrophysical discs
and internal-gravity waves in stably stratified atmospheres;\\
(ii) the presented overreflection fits in naturally within the
above-mentioned bypass concept. This allows us to adopt schemes and
ideas of the bypass concept for the understanding of driven
turbulence in MHD shear flows.

The paper is organized as follows. Section 2 is devoted to deriving
four first order differential equations (for individual
counter-propagating wave) and qualitative analysis of the equations.
Section 3 -- to the numerical analysis in two and three dimensions.
Summary and discussion are given in Section 4 where we also analyze
nonlinear consequences of the described linear dynamics and
application of the proposed approach to more complex shear flow
systems.

\section{Physical model and equations}\label{sec:equations}

Consider a 3D ideal, incompressible MHD fluid flow with
constant/linear shear of velocity, ${\bf U}_0=(Sy,0,0)$, and uniform
magnetic field, ${\bf B}_0=(B_0,0,0)$, directed along the flow. The
linearized dynamical equations for small perturbations to the flow
are:

\begin{equation}
\left( {{\partial}\over{\partial t}} + Sy {{\partial}\over{\partial
x}} \right) v_x + Sv_y = - {1 \over \rho} {{\partial
p}\over{\partial x}} ~,~~~~~~~~~~~~~~~~~~~~~\label{1}
\end{equation}
\begin{equation}
\left( {{\partial}\over{\partial t}} + Sy {{\partial}\over{\partial
x}} \right) v_y  = - {1 \over \rho} {{\partial p}\over{\partial y}}
+ {B_0 \over 4 \pi \rho}\left({{\partial b_y}\over{\partial x}} -
{{\partial b_x}\over{\partial y}}\right) ~,~\label{2}
\end{equation}
\begin{equation}
\left( {{\partial}\over{\partial t}} + Sy {{\partial}\over{\partial
x}} \right) v_z  = - {1 \over \rho} {{\partial p}\over{\partial z}}
+ {B_0 \over 4 \pi \rho} \left({{\partial b_z}\over{\partial x}} -
{{\partial b_x}\over{\partial z}}\right) ~,~ \label{3}
\end{equation}
\begin{equation}
{{\partial v_x}\over{\partial x}} + {{\partial v_y}\over{\partial
y}}+ {{\partial v_z}\over{\partial z}} = 0 ~,~~~~ {{\partial
b_x}\over{\partial x}} + {{\partial b_y}\over{\partial y}} +
{{\partial b_z}\over{\partial z}} = 0 ~,\label{4}
\end{equation}
\begin{equation}
\left( {{\partial}\over{\partial t}} + Sy {{\partial}\over{\partial
x}} \right) b_y = B_0 {{\partial v_y}\over{\partial x}}
~,~~~~~~~~~~~~~~~~~~~~~~~~~~~~~
\label{5}
\end{equation}
\begin{equation}
\left( {{\partial}\over{\partial t}} + Sy {{\partial}\over{\partial
x}} \right) b_z = B_0 {{\partial  v_z}\over{\partial x}}
~,~~~~~~~~~~~~~~~~~~~~~~~~~~~~~
\label{6}
\end{equation}
where: $\rho$ is the unperturbed density; $~p,$ $\bf v$ and $\bf b$
are the pressure, velocity and magnetic field perturbations,
respectively.

The dynamic equations permit the decomposition of perturbed
quantities into Kelvin waves, or spatial Fourier harmonics (SFHs):
\begin{equation}
\Psi(x,y,z,t)= \tilde{\Psi}(k_x,k_y(t),k_z,t) \exp
(ik_xx+ik_y(t)y+ik_zz)~,\label{7}
\end{equation}
where $~\Psi=\{p,{\bf v},{\bf b}\},$ $~\tilde{\Psi}=\{\tilde{p},{\bf
\tilde{v}},{\bf \tilde{b}}\}~$ and $~k_y(t)=k_{y0}-S k_x t.$

Introducing the following non-dimensional variables and parameters
by taking: $~1/S$ as the scale of time; Alfv${\rm \acute{e}}$n
velocity, $~V_A=B_0/(4\pi \rho)^{1/2}$, as the scale of velocity
perturbation; $~B_0$ as the scale of magnetic field perturbation,
\begin{eqnarray}
\tau=St~,~~~k_y(\tau)=k_y(0)-k_x \tau~,~~~ \hat{v}_y=\tilde{v}_y/V_A~,~~~ \nonumber\\
\hat{v}_z=\tilde{v}_z/V_A~,~~~\hat{b}_y=\tilde{b}_y/B_0~,~~~\hat{b}_z=\tilde{b}_z/B_0~,\label{8}
\end{eqnarray}
the above system of equations reduces to the following first order
dynamic equations:
\begin{equation}
{d \hat{v}_y \over d \tau}=2\chi_p(\tau) \hat{v}_y + i\Omega_{A}
\hat{b}_y ~, ~~~~~~~ {d \hat{b}_y \over d \tau}= i \Omega_{A}
\hat{v}_y ~,~~~~~~~~~\label{9}
\end{equation}
\begin{equation}
{d \hat{v}_z \over d \tau}=2\chi_s(\tau) \hat{v}_y + i\Omega_{A}
\hat{b}_z ~, ~~~~~~~ {d \hat{b}_z \over d \tau}= i \Omega_{A}
\hat{v}_z ~,~~~~~~~\label{10}
\end{equation}
where
\begin{eqnarray}
\chi_p(\tau)={k_xk_y(\tau) \over k^2(\tau)}~,~~~\chi_s(\tau)={k_xk_z
\over k^2(\tau)}~,~~~\nonumber\\
k^2(\tau) \equiv k_x^2+k_y^2(\tau)+k_z^2~,~~~\Omega_{A} \equiv
{k_xV_A \over S}~.~~~ \label{11}
\end{eqnarray}
$\Omega_{A}$ is non-dimensional Alfv${\rm \acute{e}}$n wave
frequency.

One can say, that $~\hat{v}_y$, $~\hat{v}_z$, $\hat{b}_y~$ and
$~\hat{b}_z~$ are physical variables, but not eigen ones for the
counter-propagating P-AWs and S-AWs. Here we start our optimal route
to the description of the overreflection phenomenon the essence of
which is outlined in the introduction: initially, one have to define
eigen variable for each wave components in the shearless limit (for
which the linear matrix operator is diagonal). Then, one have to
generalize the eigen variables for non-zero shear case and rewrite
the dynamic equations for them. Fortunately, in the considered MHD
flow, one can use the Els$\ddot{a}$sser variables as the generalized
ones,
\begin{eqnarray}
Z_p^{\pm}= \hat{v}_y \mp \hat{b}_y~,~~~Z_s^{\pm}= \hat{v}_z \mp
\hat{b}_z~.~~~\label{12}
\end{eqnarray}

Inserting the Els$\ddot{a}$sser variables into equations (2.9)
and(2.10), one get the set of the four, first order differential
equations describing the dynamics of P-Alfv${\rm \acute{e}}$n
(labelled by the index ``p'') and S-Alfv${\rm \acute{e}}$n (labelled
by the index ``s'') wave SFHs:
\begin{equation}
{d Z_p^{+} \over d \tau} = - i \Omega_{A} Z_p^{+} + \chi_p(\tau)
Z_p^{+} + \chi_p(\tau) Z_p^{-}~,~~~\label{13}
\end{equation}
\begin{equation}
{d Z_p^{-} \over d \tau} = i\Omega_{A} Z_p^{-} +  \chi_p(\tau)
Z_p^{-} + \chi_p(\tau) Z_p^{+}~,~~~~ ~\label{14}
\end{equation}
\begin{equation}
{d Z_s^{+} \over d \tau} = - i \Omega_{A} Z_s^{+} + \chi_s(\tau)
Z_p^{+} + \chi_s(\tau) Z_p^{-}~,~~~\label{15}
\end{equation}
\begin{equation}
{d Z_s^{-} \over d \tau} = i\Omega_{A} Z_s^{-} +  \chi_s(\tau)
Z_p^{-} + \chi_s(\tau) Z_p^{+}~.~~~~~\label{16}
\end{equation}

One can see that the linear matrix operator of the equations
(written for the eigen variables) is not diagonal in shear flow.
I.e. the dynamics of the waves are not separated to each other any
more.

\subsection{Spectral energy}

Strictly speaking, the energy of waves should be determined in the
framework of nonlinear problems. However, usually, the concept of
energy is introduced in solving linearized problems
\citep{Stepanyants89}. We also do the same to get a feeling of the
dynamics of quadratic forms of physical variables.

The energy of an individual SFH (total spectral energy) is the sum
of kinetic and magnetic ones:
\begin{eqnarray}
E(k_x,k_y(\tau),\tau) ={\rho V_A^2 \over 2} \left(|\hat{v}_x|^2 +
|\hat{v}_y|^2
+|\hat{v}_z|^2 + |\hat{b}_x|^2 + |\hat{b}_y|^2  + |\hat{b}_z|^2\right) = ~~~~~~~~~~~~~~~~~~~~~~~~~~~~~~~ \nonumber\\
= {1 \over 4}\rho V_A^2 \left(1 + {{k_y^2(\tau)} \over {k_x^2}}
\right )\left(|Z_p^{+}|^2 + |Z_p^{-}|^2\right) + {1 \over 4}\rho
V_A^2 \left(1 + {{k_z^2} \over {k_x^2}}
\right )\left(|Z_s^{+}|^2 + |Z_s^{-}|^2\right) + ~~~~~~~~~~~~~~~~~~~~~\nonumber\\
+ {1 \over 4}\rho V_A^2 {{k_y(\tau)k_z} \over {k_x^2}}
\left(Z_p^{+}Z_s^{+*} + Z_p^{-}Z_s^{-*}+Z_s^{+}Z_p^{+*} +
Z_s^{-}Z_p^{-*}\right) =
~~~~~~~~~~~~~~~~~~~~~~~~~~\nonumber\\
= E_p^{+} + E_p^{-} + E_s^{+} + E_s^{-} + E_{int},~~~~~~ \label{17}
\end{eqnarray}
where:
\[
E_i^{\pm} \equiv
 {1 \over 4}\rho V_A^2 \left(1 + {{k_y^2(\tau)} \over {k_x^2}}
\right )|Z_i^{\pm}|^2 , ~~~ i=p,s~~~~~~~~~~~~~~~~~~~~~~~~~~~~
\]
\[
E_{int} \equiv {1 \over 4}\rho V_A^2 {{k_y(\tau)k_z} \over {k_x^2}}
\left(Z_p^{+}Z_s^{+*} + Z_p^{-}Z_s^{-*}+Z_s^{+}Z_p^{+*} +
Z_s^{-}Z_p^{-*}\right).
\]

One can say, that $E_p=E_p^{+} + E_p^{-}$ is the spectral energy of
P-AWs, $E_s=E_s^{+} + E_s^{-}$ -- the spectral energy of S-AWs, and
$E_{int}$ -- the spectral potential energy connected to the
coupling, or interaction between P-AWs and S-AWs. So, the spectral
energy, that is the sum of quadratic forms of velocity and magnetic
field perturbations, does not reduce to a sum of quadrates of the
Els$\ddot{a}$sser variables -- it can be interpreted as the sum of
the spectral energies of P-AWs and S-AWs with the additional term,
$E_{int}$. This additional term appears in the framework of the
considered linear theory due to the coupling of these two waves and,
of course, has not any relation to their nonlinear interaction.
Generally, $E_{int}$ is of the order of $E_p$ and $E_s$ (see below
figure \ref{fig:fig.6}). Evaluating the dynamic processes, we will
work in terms of the spectral energies of the waves ($E_p$, $E_s$,
$E_p^{+}$, $E_s^{+}$, $E_p^{-}$ and $E_s^{-}$).

\subsection{Qualitative analysis of the dynamic equations: the wave linear coupling as the basis of the overreflection}

From equations (2.13)-(2.16) it follows that the dynamics of
counter-propagating P-AW SFHs is self-contained -- the dynamics is
defined by the intrinsic to the P-AWs terms, while the dynamics of
S-AW SFHs is defined by the extrinsic to S-AWs terms -- the second
and third rhs terms of equations (2.15) and (2.16) linearly couple
the dynamics of each S-AW SFH to the corresponding SFH of P-AWs. So,
the coupling is asymmetric: S-AWs do not participate in the dynamics
of P-AWs, while P-AWs do. Somewhat similar investigation, but in
compressible case, is performed in \citet{Hollweg12}. In
compressible flows, the coupling is mutual -- slow magnetosonic
waves are generated by Alfv${\rm \acute{e}}$n ones and inverse. A
physical result of this process is the generation of density
perturbations by Alfv${\rm \acute{e}}$n waves.

The dynamics of each (counter-propagating) P-AW SFH is determined by
the interplay of three different terms on the right hand side of
equations (2.13) and (2.14). The second rhs terms of these equations
relate to a mechanism of energy exchange between the mean flow and
the SFH. The third rhs terms couple these equations, or physically,
linearly couple the counter-propagating P-AWs. These terms relate to
another mechanism, which is responsible, in many respects, for P-AW
overreflection phenomenon. $\chi_p$ is the time-dependent coupling
coefficient of counter-propagating P-AWs and, in fact, its value
determines the strength of the overreflection of these waves. In the
shearless limit, $\chi_p=0$ and only the first, easily recognizable
terms are left in these equations, which result in oscillations with
normalized frequencies $~\Omega_{A}$ and $-\Omega_{A}$, i.e., with
Alfv${\rm \acute{e}}$n frequency, $K_xV_A$, in dimensional
variables.

As for equations (2.15) and (2.16), the second and third rhs terms
describe the growth of S-AWs that occurs due to the linear coupling
of P-AWs and S-AWs, i.e. the growth of S-AWs is an indirect
consequence of P-AWs growth. Nevertheless, as it follows from the
below performed numerical calculations, the growth of P-AWs prevails
over the growth of S-AWs for a wide range of the system parameters.

\subsection{Qualitative analysis of the dynamics of P-Alfv${\rm \acute{e}}$n waves}

We analyze the wave dynamics in polar coordinates,
\begin{equation}
Z_i^{\pm}(\tau)=|Z_i^{\pm}(\tau)|\exp(-i\phi_i^{\pm}(\tau))~,~~~
i=p,s~~~
\label{18}
\end{equation}
and define the degree of imbalance for counter-propagating SFHs of
P-AWs and S-AWs as:
\begin{equation}
\alpha_i = 1 - {|Z_i^{-}|^2 \over|Z_i^{+}|^2}~,~~~
i=p,s.~~~
\label{19}
\end{equation}

For the purpose of the visualization of overreflection phenomenon we
introduce ``instantaneous frequency'' as
\begin{equation}
\Omega_{i}^{\pm}(\tau)= {d {\phi_i^{\pm}(\tau)} \over {d
\tau}}~,~~~i=p,s.~~~
\label{20}
\end{equation}

As it was mentioned above, the dynamics of counter-propagating
P-Alfv${\rm \acute{e}}$n waves is self-contained. Let's focus on a
qualitative analysis of P-AWs' SFHs amplitude and phase dynamics.

From equations (2.13) and (2.14) it follows:
\begin{eqnarray}
{{d }\over {d \tau}} (|Z_p^{+}|^2- |Z_p^{-}|^2) =
2\chi_p(\tau)(|Z_p^{+}|^2 - |Z_p^{-}|^2) =
~~~~~~~~~~~~~~~~~~~~~~~~~~~~~~~~~\nonumber\\
=\left[{{d }\over {d \tau}}\ln \left({{k_x^2 + k_y^2(0)+
k_z^2}\over{k_x^2 + k_y^2(\tau)}+ k_z^2} \right )\right
](|Z_p^{+}|^2 - |Z_p^{-}|^2) ~, \label{21}
\end{eqnarray}
or, after integration,
\begin{eqnarray}
|Z_p^{+}(\tau)|^2 - |Z_p^{-}(\tau)|^2 = {{k_x^2 + k_y^2(0) +
k_z^2}\over{k_x^2 + k_y^2(\tau)+ k_z^2}} (|Z_p^{+}(0)|^2 -
|Z_p^{-}(0)|^2)~.\label{22}
\end{eqnarray}

We see that if the waves are balanced at the beginning,
$|Z_p^{+}(0)| = |Z_p^{-}(0)|$, they remain balanced. If at the
beginning $|Z_p^{+}(0)| \neq |Z_p^{-}(0)|$, then the difference of
the intensities varies by an algebraic law ($\sim 1/k^2(\tau)$) that
is characteristic to transient growth in hydrodynamic shear flows
\citep{Farrell93b,Schmid07}. This indicates a common basis of
transient dynamics in MHD and hydrodynamics shear flows.

Renormalizing the Els$\ddot{a}$sser variables:
\begin{eqnarray}
|Z_p^{\pm}(\tau)|= \exp\left(\int \limits_{0}^{\tau} d \tau^{\prime}
\chi_p(\tau^{\prime})\right)|{\hat Z}_p^{\pm}(\tau)| =\sqrt {{k_x^2
+ k_y^2(0) + k_z^2}\over{k_x^2 + k_y^2(\tau)+ k_z^2}}|{\hat
Z}_p^{\pm}(\tau)|~, \label{23}
\end{eqnarray}

Equations (2.13) and (2.14) are reduced to
\begin{equation}
{d {\hat Z}_p^{+} \over d \tau} = - i \Omega_{A} {\hat Z}_p^{+} +
\chi_p(\tau) {\hat Z}_p^{-}~,~~~ {d {\hat Z}_p^{-} \over d \tau} =
i\Omega_{A} {\hat Z}_p^{-} + \chi_p(\tau) {\hat Z}_p^{+}~,~~~~~
\label{24}
\end{equation}
and equation (2.22) to
\begin{equation}
|{\hat Z}_p^{+}(\tau)|^2 - |{\hat Z}_p^{-}(\tau)|^2 = |{\hat
Z}_p^{+}(0)|^2 - |{\hat Z}_p^{-}(0)|^2~, \label{25}
\end{equation}
i.e., in this version of eigen variables, the dynamic equations are
simplified (compare 2.13 and 2.14 with 2.24) and, in addition (as it
follows from equation 2.25), the difference between the intensities
is constant. Equation (2.25) indicates the conservation of action
for P-AW harmonics. The similar conservation of wave action for
different and more complex configuration is derived in
\citet{Heinemann80}. The latter considers small-amplitude, toroidal
non-WKB (long wavelength) Alfv${\rm \acute{e}}$n waves in a model of
axisymmetric ideal MHD solar wind flow neglecting solar rotation.
The considered toroidal waves decouple from compressional waves in
linear approximation and their amplitudes dynamics can be computed
from only two equations without consideration of the other wave
modes. I.e. the dynamics of the toroidal Alfv${\rm \acute{e}}$n
waves is self-contained as the dynamics of counter-propagating P-AWs
considered here. Consequently, the conservation of wave action in
the both cases is reduced to the conservation of the wave action for
one (counter-propagating, or inward-outward) wave mode independently
to the complexity of the flow system and has the simple form.

Equations (2.18) and (2.23) give
\begin{equation}
{\hat Z}_p^{\pm}(\tau)=|{\hat
Z}_p^{\pm}(\tau)|\exp(-i\phi_p^{\pm}(\tau))~,~~~
\label{27}
\end{equation}
substituting which in (2.24), after simple but cumbersome
mathematical manipulations, finally, results in two dynamic
equations for the normalized total intensity, $|{\hat
Z}_p^{+}(\tau)|^2+|{\hat Z}_p^{-}(\tau)|^2$, and phases difference,
$\Delta \phi_p(\tau)=\phi_p^{+}(\tau) - \phi_p^{-}(\tau)$, of the
counter-propagating P-Alfv${\rm \acute{e}}$n waves:
\begin{equation}
{d \ln (|{\hat Z}_p^{+}|^2+|{\hat Z}_p^{-}|^2) \over d \tau}
=\Gamma(\tau) \chi_p(\tau) \cos \Delta \phi_p ~, \label{28}
\end{equation}
\begin{equation}
{d \Delta \phi_p \over d \tau} =2\Omega_A - {\chi_p(\tau) \over
\Gamma(\tau)} \sin \Delta \phi_p ~, \label{29}
\end{equation}
where,
\begin{equation}
\Gamma(\tau)={{2|{\hat Z}_p^{+}||{\hat Z}_p^{-}|}\over{|{\hat
Z}_p^{+}|^2+|{\hat Z}_p^{-}|^2}}~ \label{30}
\end{equation}
mathematically is the ratio of the geometrical and arithmetic means
of the amplitudes. Of course, this ratio is the maximum when the
amplitudes are equal to each other. Consequently, the fastest growth
of the total intensity of the counter-propagating waves occurs when
the waves are balanced from the beginning, $|{\hat Z}_p^{+}(0)| =
|{\hat Z}_p^{-}(0)|$. In this case, $\Gamma(\tau) = 1$ that
contributes to the intensification of the growth. The results of the
presented below numerical calculations confirms this.

The growth also depends on sign-varying quantities $~\chi_p$ and
$~\cos \Delta \phi_p$. For the optimal growth, the coincidence of
their signs during main part of the dynamics is necessary. The sign
of $~\chi_p(\tau)$ is defined by the sign of $~k_xk_y(\tau)$. If
initially $~k_xk_y(0)>0$, $~\chi_p(0)$ is positive. In the course of
time, when $~\tau > \tau^*\equiv k_y(0)/k_x$, $\chi_p(\tau)$ becomes
negative. For the effectivness of the growth, well-timed change of
the sign of $~\cos \Delta \phi_p$ is necessary to make the rhs of
equation (2.27) positive again. So, the growth should depend
strongly on the dynamics of $~\Delta \phi_p(\tau)$, including the
initial value of the phase difference, $~\Delta \phi_p(0)$. This
fact is also confirmed by the following calculations.

\section{Numerical analysis}

It is seen from equations (2.13)-(2.16) that, S-AWs do not
participate in any energy exchange processes in the flow. If
initially only S-AWs are excited in the flow, $|Z_s^{\pm}(0)|\neq 0$
and $|Z_p^{\pm}(0)|=0$, and the dynamics is trivial -- any kind of
energy exchange process is absent and we simply have the propagation
of S-AWs. So, we analyze cases when initially only P-AW SFHs are
imposed in the flow. Specifically, we inserted a single
unidirectional P-AW harmonic (i.e., with one sign of frequency,
$Z_p^{+}(0)=1$ and $Z_p^{-}(0)= 0$) or counter-propagating P-AW
harmonics with equal amplitudes but different phases ($Z_p^{+}(0)=1$
and $Z_p^{-}(0)= 1,i,-1$). We present the results of the numerical
calculations for $\Omega_{A} = 0.1;0.3;1$ and $k_y(0)/k_x=100$. We
consider two-dimensional $(k_z=0)$, as well as three-dimensional
$(k_z/k_x=1;10)$ cases. A general outcome of the dynamics is the
following: the growth of the waves occurs mostly at $\Omega_{A} < 1$
and $k_y(0)/k_x > 1$; the intensity of the processes increases with
the decrease of $\Omega_{A}$ and increase of $k_y(0)/k_x$; it also
strongly depends on the value of $k_z/k_x$.

\subsection{Two-dimensional case}

\begin{figure}
  \centerline{\includegraphics[width=9cm]{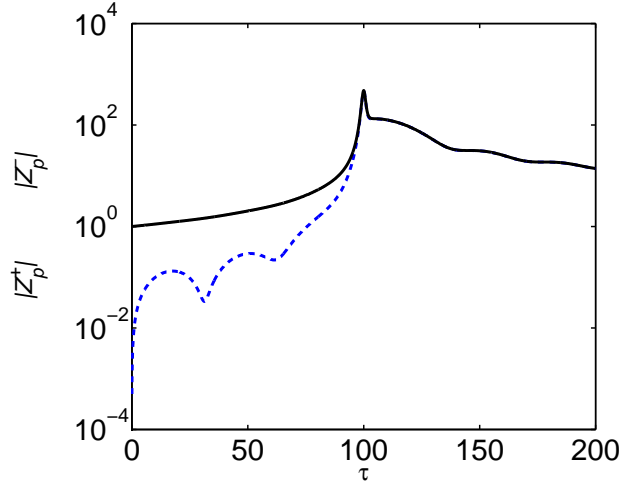}}
\caption{$~|Z_p^{+}|$ (solid  black) and $~|Z_p^{-}|$ (dashed blue)
vs time in log-linear scaling at: $Z_p^{+}(0)=1$, $Z_p^{-}(0)=0$,
$Z_s^{+}(0)=0$, $Z_s^{-}(0)=0$, $\Omega_{A}=0.1$, $~k_y(0)/k_x=100$
and  $k_z=0$.}
\label{fig:fig.1}
\end{figure}

\begin{figure}
    \centerline{\includegraphics[width=8.5cm]{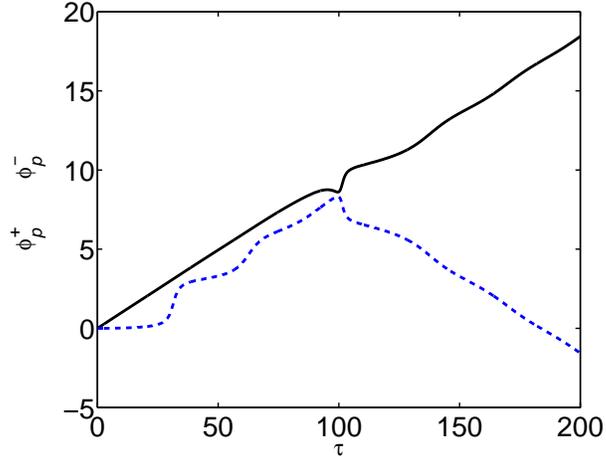}}
\caption{$~\phi_p^{+}$ (solid  black) and $~\phi_p^{-}$ (dashed
blue) vs time for the same parameters as in figure \ref{fig:fig.1}.}
\label{fig:fig.2}
\end{figure}

\begin{figure}
    \centerline{\includegraphics[width=9cm]{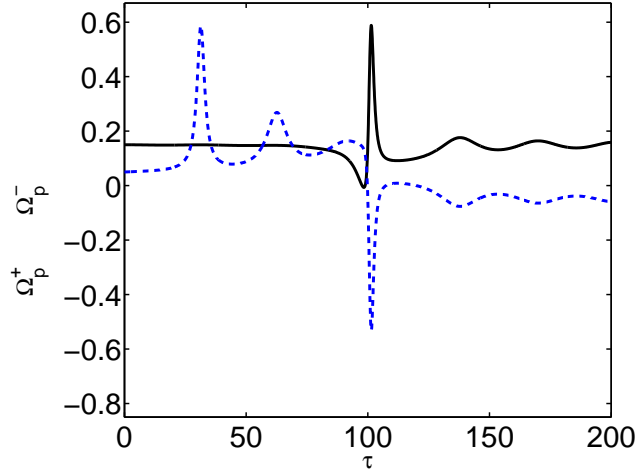}}
\caption{Instantaneous frequencies $\Omega_{p}^{+}$ (solid  black)
and $~\Omega_{p}^{-}$ (dashed blue) vs time for the same parameters
as in figure \ref{fig:fig.1}.}
\label{fig:fig.3}
\end{figure}

The transient dynamics of amplitudes of P-AWs (governed by equations
2.13 and 2.14) is shown in figure \ref{fig:fig.1} at $\Omega_{A} =
0.1$ and $k_z=0$ when initially only unidirectional P-AW harmonic is
imposed in the flow. The dynamics of corresponding phases of the
two-dimensional P-AW harmonic is presented in figure
\ref{fig:fig.2}. Plotted in figure \ref{fig:fig.3} are the variation
of the P-AWs' SFH instantaneous frequencies, $\Omega_{p}^{+}$ and
$\Omega_{p}^{-}$, associated with the dynamics of the corresponding
phases.

\begin{figure}
    \centerline{\includegraphics[width=9cm]{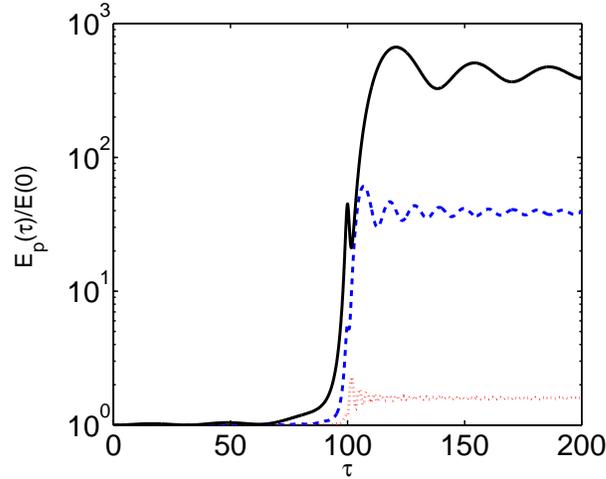}}
\caption{Normalized spectral energy of P-AWs, $E_p(\tau)/E(0)$, vs
time in log-linear scaling for the same parameters as in figure
\ref{fig:fig.1}, but at $\Omega_{A}=0.1$ (solid  black),
$~\Omega_{A}=0.3$ (dashed blue) and  $~\Omega_{A}=1$ (dotted red).}
\label{fig:fig.4}
\end{figure}

\begin{figure}
    \centerline{\includegraphics[width=8.5cm]{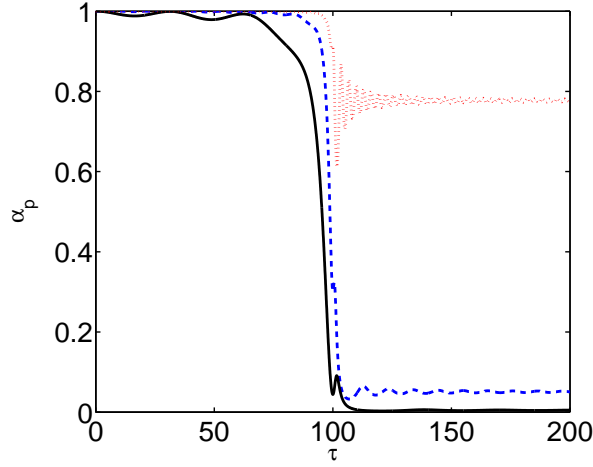}}
\caption{The imbalance degree of P-AWs, $\alpha_p = 1 - |Z_p^{-}|^2
|Z_p^{+}|^2$, vs time for the same parameters as in figure
\ref{fig:fig.1}, but at $~\Omega_{A}=0.1$ (solid  black),
$~\Omega_{A}=0.3$ (dashed blue) and  $~\Omega_{A}=1$ (dotted red).}
\label{fig:fig.5}
\end{figure}

\begin{figure}
\centerline{\includegraphics[width=8.5cm]{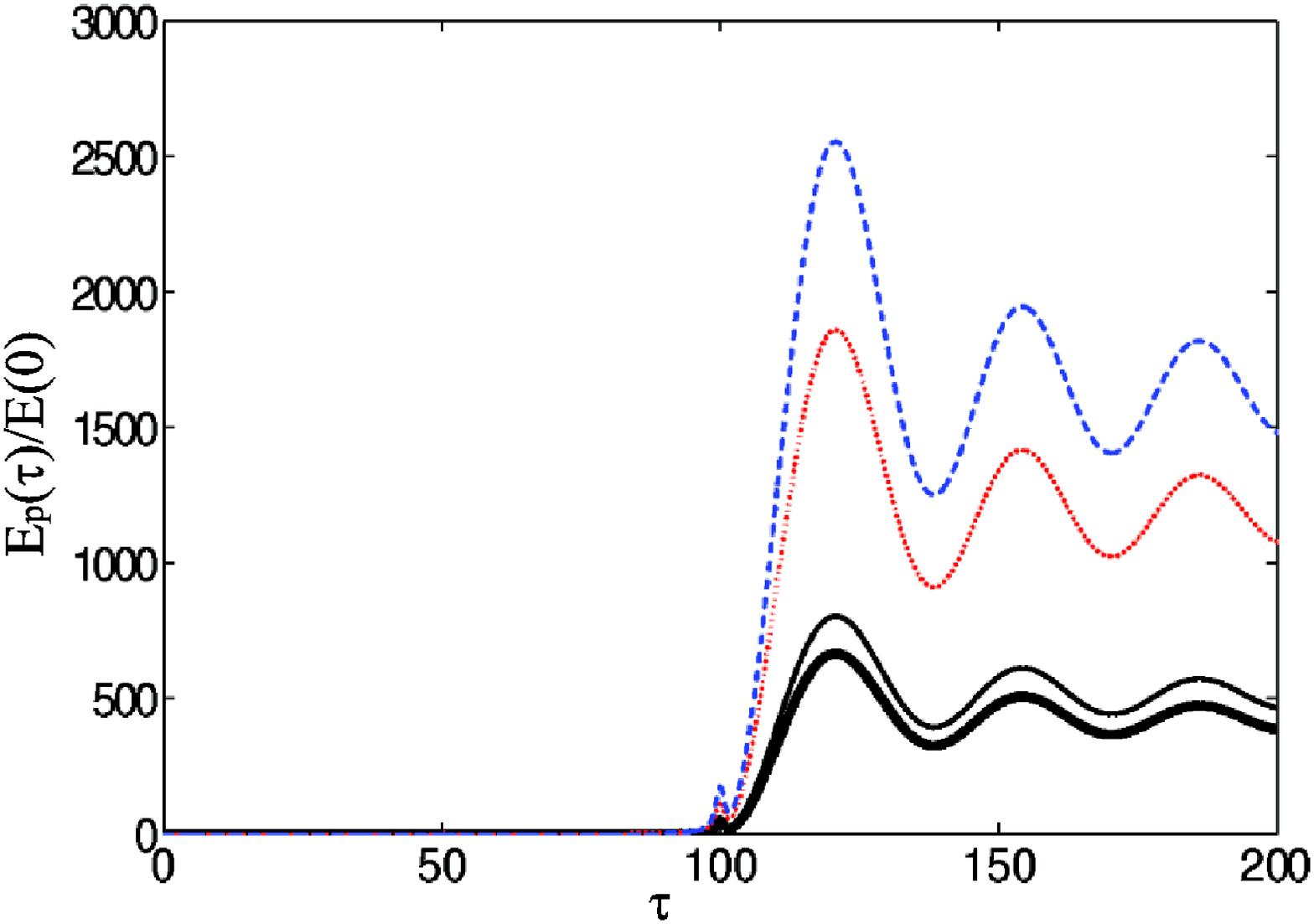}}
\caption{Normalized spectral energy of P-AWs, $E_p(\tau)/E(0)$, vs
time in two-dimensional case ($k_z=0$, $Z_s^{\pm}=0$) at:
$\Omega_{A}=0.1$, $~k_y(0)/k_x=100$, $Z_p^{+}(0)=1$ and
$Z_p^{-}(0)=0$ (thick solid black), $Z_p^{-}(0)=1$ (solid black),
$Z_p^{-}(0)=i$ (dashed blue), $Z_p^{-}(0)=-1$ (dotted red).}
\label{fig:fig.6}
\end{figure}

\begin{figure}
\centerline{\includegraphics[width=9cm]{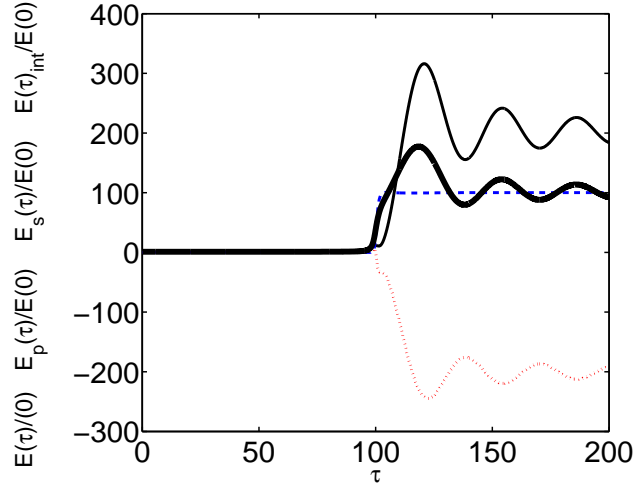}}
\caption{Normalized spectral energies:  $E(\tau)/E(0)$ (thick solid
black), $E_p(\tau)/E(0)$ (solid  black),  $E_s(\tau)/E(0)$ (dashed
blue) and  $E_{int}(\tau)/E(0)$ (dotted red) vs time at:
$Z_p^{+}(0)=1$, $Z_p^{-}(0)=0$, $Z_s^{\pm}(0)=0$, $\Omega_{A}=0.1$,
$~k_y(0)/k_x=100$ and  $k_z/k_x=1$.}
\label{fig:fig.7}
\end{figure}

\begin{figure}
\centerline{\includegraphics[width=8.5cm]{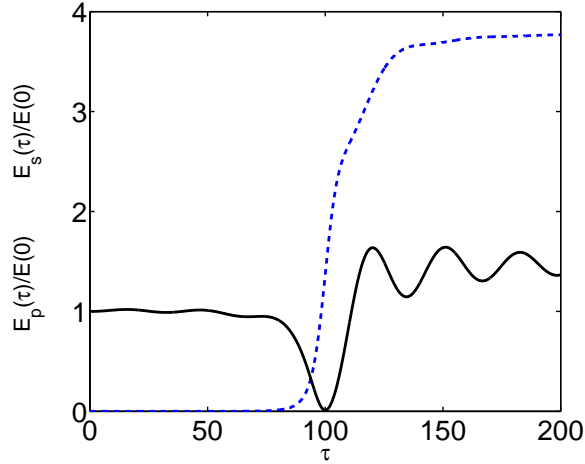}}
\caption{Normalized spectral energies of P-AWs and S-AWs,
$E_p(\tau)/E(0)$ (solid  black) and  $E_s(\tau)/E(0)$ (dashed blue),
vs time for the same parameters as in figure \ref{fig:fig.7}, but at
$k_z/k_x=10$.} \label{fig:fig.8}
\end{figure}

\begin{figure}
\centerline{\includegraphics[width=8.5cm]{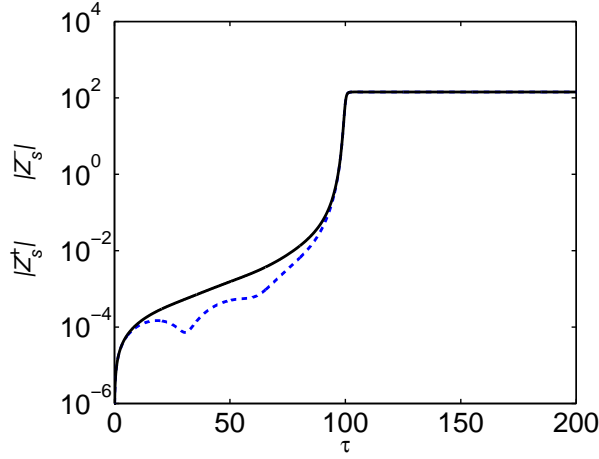}}
\caption{$~|Z_s^{+}|$ (solid  black) and $~|Z_s^{-}|$ (dashed blue)
vs time in log-linear scaling for the same parameters as in figure
\ref{fig:fig.7}.} \label{fig:fig.9}
\end{figure}

\begin{figure}
\centerline{\includegraphics[width=8.5cm]{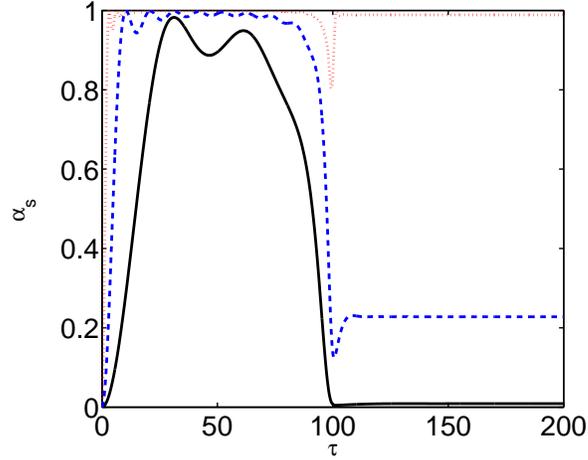}} \caption{The
imbalance degree of S-AWs, $\alpha_s = 1 - |Z_s^{-}|^2
/|Z_s^{+}|^2$, vs time in log-linear scaling for the same parameters
as in figure \ref{fig:fig.7}, but at $~\Omega_{A}=0.1$ (solid
black), $~\Omega_{A}=0.3$ (dashed blue) and  $~\Omega_{A}=1$ (dotted
red).} \label{fig:fig.10}
\end{figure}

\begin{figure}
\centerline{\includegraphics[width=7cm]{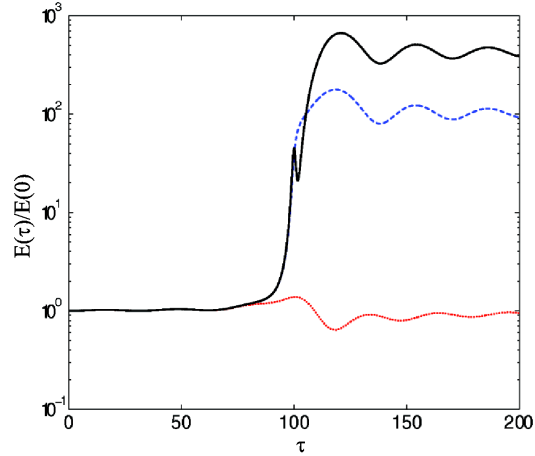}}
\caption{Normalized spectral energy, $E(\tau)/E(0)$, vs time in
log-linear scaling for 2D and 3D cases for the same parameters as in
figure \ref{fig:fig.7}, but at $k_z/k_x=0$ (solid),  $k_z/k_x=1$
(dashed) and $k_z/k_x=10$ (dotted).} \label{fig:fig.11}
\end{figure}

\begin{figure}
\centerline{\includegraphics[width=7cm]{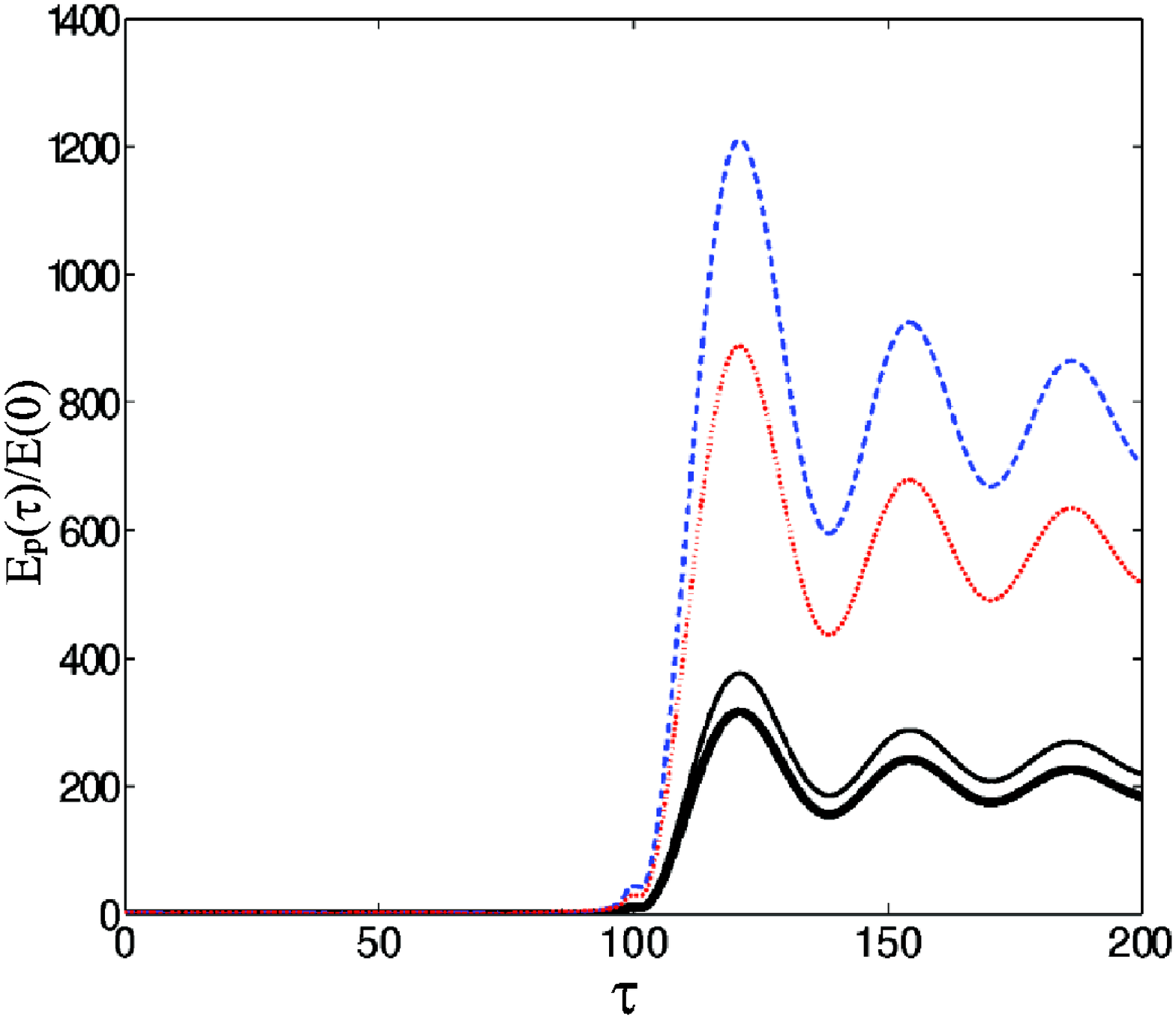}}
\caption{Normalized spectral energy of P-AWs, $E_p(\tau)/E(0)$, vs
time at: $\Omega_{A}=0.1$, $~k_y(0)/k_x=100$, $k_z/k_x=1$,
$Z_s^{\pm}(0)=0$, $Z_p^{+}(0)=1$ and $Z_p^{-}(0)=0$ (thick solid
black), $Z_p^{-}(0)=1$ (solid black), $Z_p^{-}(0)=i$ (dashed blue),
$Z_p^{-}(0)=-1$ (dotted red).} \label{fig:fig.12}
\end{figure}

\begin{figure}
\centerline{\includegraphics[width=7cm]{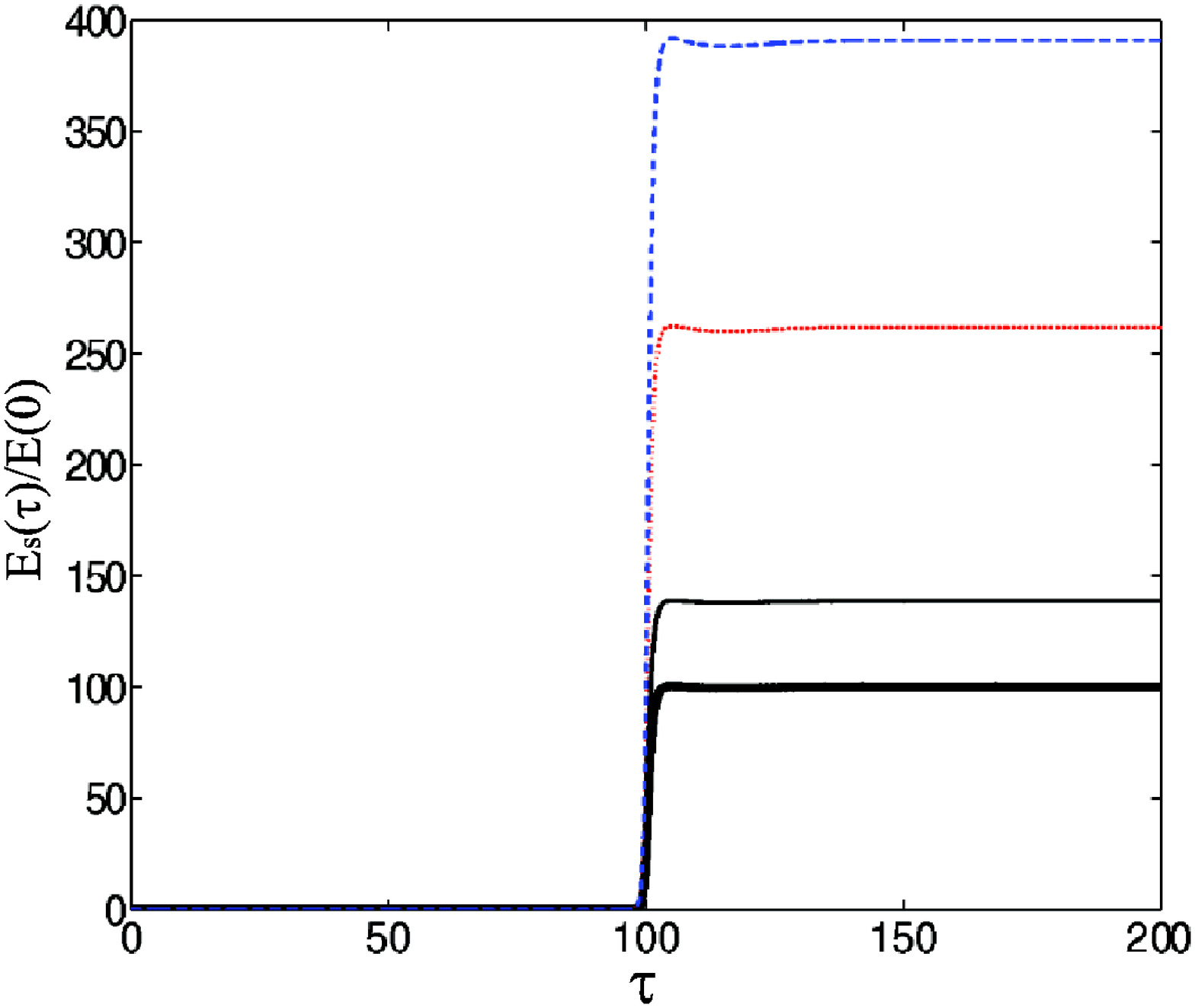}}
\caption{Normalized spectral energy of S-AWs, $E_s(\tau)/E(0)$, vs
time at: $\Omega_{A}=0.1$, $~k_y(0)/k_x=100$, $k_z/k_x=1$,
$Z_s^{\pm}(0)=0$, $Z_p^{+}(0)=1$ and  $Z_p^{-}(0)=0$ (thick solid
black), $Z_p^{-}(0)=1$ (solid black), $Z_p^{-}(0)=i$ (dashed blue),
$Z_p^{-}(0)=-1$ (dotted red).} \label{fig:fig.13}
\end{figure}

With the help of equations (2.13) and (2.14) and Figures
\ref{fig:fig.1}-\ref{fig:fig.3} one can trace each stage of the
evolution of the counter-propagating P-AW harmonic, which in fact,
represents the overreflection phenomenon. Initially, as
$Z_p^{-}(0)=0$, in (2.14) only the last rhs term is nonzero. So, the
initial amplification and the dynamics of $Z_p^{-}(\tau)$ is due to
the third term $\chi_p Z^{+}$. Therefore, the positive
``instantaneous frequency'' of $Z_p^{+}$ results in the positive
``instantaneous frequency'' of $Z_p^{-}$ (see figure
\ref{fig:fig.3}). The growth of $|Z_p^{-}|$ is rapid, but algebraic
(nonexponential). In the course of the evolution, $|Z_p^{-}|$
becomes almost equal to $|Z_p^{+}|$ (see figure \ref{fig:fig.1} at
$\tau \simeq \tau^*\equiv k_y(0)/k_x=100$). At the same time, the
influence of the first and second rhs terms of equation (2.14)
become appreciable, behavior of $\phi_p^{-}$ and $\Omega_{p}^{-}$
changes at $\tau \simeq \tau^*$ with $\Omega_{p}^{-}$ becoming
negative and, as a result of all these, $Z_p^{-}$ is propagating
opposite to $Z_p^{+}$. With further increase of time,
$\Omega_{p}^{-}$ tends to $-\Omega_{A}$ and slightly varies around
it. As for the dynamics of $Z_p^{+}$, the coupling (the last rhs
term of equation 2.13) somewhat modifies its dynamics in the
vicinity of $\tau \simeq \tau^*$, where $Z_p^{-}$ is already large
and $\chi_p$ is not small too (while at $\tau \gg \tau^*,~\chi_p
\rightarrow 0$).

Figure \ref{fig:fig.4} shows that, starting with a purely
unidirectional P-AW harmonic, $Z_p^{-}(0)=0$, the perturbation
energy increases and reaches a peak value at time $\tau \simeq
\tau^*$. After that, the harmonic undergoes nearly periodic and
damping oscillations around some plateau value of the energy. This
plateau value increases with decreasing $\Omega_{A}$.

Figure \ref{fig:fig.5} shows that the imbalance degree of the P-AW
harmonic decreases with time (with the increase of the amplitudes),
i.e., the energy propagating opposite to the $X$-axis approaches the
energy propagating along the $X$-axis: for the case of
$\Omega_{A}=1$, the harmonic remains imbalanced, for the case of
$\Omega_{A}=0.3$ the imbalance degree tends to $0.05$, and for the
case of $\Omega_{A}=0.1$, in fact, the harmonic becomes balanced
with time.

When initially counter-propagating P-AW harmonics with equal
amplitudes but different phases ($Z_p^{+}(0)=1$ and $Z_p^{-}(0)=
1,i,-1$) are inserted, the dynamics is simpler. Amplitudes of the
physical variables are equal to each other,
$Z_p^{+}(\tau)=Z_p^{-}(\tau)$, according to equation (2.22), i.e.,
they are balanced from the beginning and their phase dynamics is
quite trivial.

Figure \ref{fig:fig.6} shows that the transient growth of
perturbation energy is smaller in the case of initially imposed
unidirectional P-AW harmonic than that in the case when equal
amplitude counter-propagating P-AW harmonics are imposed. And, in
the latter cases, the energy growth is maximum if the phases of the
inserted harmonics differ from each other by $\pi/2~$
($Z_p^{+}(0)=1$ and $Z_p^{-}(0)= i$).

\subsection{Three-dimensional case}

Figures \ref{fig:fig.7}-\ref{fig:fig.11} correspond to 3D cases when
initially just unidirectional P-AW harmonic is imposed in the flow.
Figures \ref{fig:fig.12} and \ref{fig:fig.13} -- when initially
counter-propagating P-AW harmonics are inserted with equal
amplitudes but different phases.

In the 3D case, S-AWs become an active participant in the dynamics,
however, the growth of P-AWs remains somewhat larger than the growth
of S-AWs. Figure \ref{fig:fig.7} shows the prevalence of P-AWs for
$k_z/k_x = 1$. (It is natural that this prevalence is more
pronounced for $~k_z/k_x \ll 1$.) The transient growth of both waves
substantially reduces with the further increase of the ratio
$~k_z/k_x$ (see figure \ref{fig:fig.8}) and, consequently, SFHs with
$~k_z/k_x \gg 1$ do not play any role in the dynamical processes.

In figure \ref{fig:fig.9} we present the dynamics of $|Z_s^{+}|$
(solid black) and $|Z_s^{-}|$ (dashed blue) vs time in log-linear
scaling for a small value of $\Omega_{A}$ ($\Omega_{A}=0.1$) and
$k_z/k_x=1$ when, only $Z_p^{+}(0)$ is inserted in the flow (i.e.,
$Z_p^{-}(0),Z_s^{\pm}(0)=0$). In the beginning, $|Z_s^{+}|$
increases stronger than $|Z_s^{-}|$. However, in the course of the
evolution, $|Z_s^{-}|$ becomes almost equal to $|Z_s^{+}|$. Figure
\ref{fig:fig.10} shows that with increase of the amplitudes, the
imbalance degree of the S-AWs decreases with time for small
$\Omega_{A}$ as it is for P-AWs (compare figures \ref{fig:fig.5} and
\ref{fig:fig.10}). S-AWs are imbalanced already at $\Omega_{A}\simeq
1$. However, the growth of S-AWs is negligible in the last case.

Figure \ref{fig:fig.11} shows that, the growth of the total spectral
energy is maximal for 2D case and decreases with the increase of
$~k_z/k_x$. This result is somewhat unexpected/surprising, because
in non-magnetized flows (i.e., in the simplest incompressible
hydrodynamic constant shear flow) the transient growth of
three-dimensional perturbations is generally stronger than transient
growth of two-dimensional ones \citep[e.g., see][]{Chagelishvili96,
Moffat67, Farrell93b, Bakas01} and the dynamics of a non-magnetized
system is determined by three-dimensional perturbations.

Figures \ref{fig:fig.12} and \ref{fig:fig.13} show that the energy
growth is maximal if the phases of inserted SFHs differ from each
other by $\pi/2$ (as in the 2D case). These figures also coincide
with \ref{fig:fig.7} for cases when phases of the initially inserted
SFHs differ from each other -- the transient growth of P-AW always
prevails over the growth of S-AW. All energy dynamics plots (see
figures
\ref{fig:fig.4},\ref{fig:fig.6},\ref{fig:fig.7},\ref{fig:fig.11}-\ref{fig:fig.13})
show that real transient growth takes place in the vicinity of $\tau
\simeq \tau^*$ during the time interval $|\tau - \tau^*| \leq
\Omega_A^{-1}$.

\section{Summary and discussion}

The proposed formalism provides a deeper insight into the physical
mechanism underlying the overreflection phenomenon by allowing us to
separate from each other physical processes associated with
counter-propagating waves and to follow their interaction during the
overreflection. Based on the presented analysis, the path to the
overreflection is as follows. Initially an imposed on the shear flow
pure $Z_p^{+}$ generates $Z_p^{-}$ due to the shear-induced linear
coupling, that at first propagates in the same direction as
$Z_p^{+}$. With time, $Z_p^{-}$ grows transiently and in the
vicinity of $\tau \simeq \tau^*$ (at which $~k_y(\tau^*)=0$)
reverses the direction of propagation and becomes
counter-propagating to $Z_p^{+}$. Both counter-propagating P-AWs
SFHs exhibit transient growth that is appreciable at $\Omega_{A} <
1$, $k_z/k_x \leq 1$ and $k_y(0)/k_x \gg 1$. As for S-AWs, its
transient growth occurs due to the linear coupling of P-AW and S-AW,
i.e. -- is an indirect consequence of P-AW (i.e. $Z_p^{+}$ and
$Z_p^{-}$) growth. At the same time, the growth of P-AWs somewhat
prevails over the growth of S-AWs. It is obvious, that the dynamics
has transient nature, as it should be -- shear flow non-normality
induced energy exchange processes are always transient \citep[e.g.,
see][]{Schmid07}. This statement is correct for constant in time
shear flows. However, in periodic shear flows (i.e., when shear
parameter is a periodic function of time), wave perturbations may
grow exponentially in time. For instance,
\cite{Zaqarashvili00,Zaqarashvili02} studied the stability of
periodic MHD shear flows, showing that the temporal behaviour of
spatial Fourier harmonics of magnetosonic waves is governed by
Mathieu's equation. Consequently, the harmonics with the half
frequency of the shear flow grow exponentially in time. Mathieu's
equation represent a single second order ordinary differential
(wave) equation and describes counter-propagating waves of a
single/physical variable. Consequently, possible dynamical processes
between these waves (e.g., their coupling) are, in fact, left out of
consideration. To get an additional information about the physics of
the growth (e.g., to grasp the coupling between counter-propagating
waves) one has to find ``eigen-variable'' for each wave component
and, by this way, reduce Mathieu's equation to the set of first
order differential equations for individual counter-propagating wave
(as it is done for the constant shear flows).

Evaluating the linear transient dynamics we use the concept of
energy. Of course, when calculating energy - the quantity of the
second order in the wave amplitude - causes dissatisfaction
associated with the known fact: it is obviously necessary to take
into account changes in the mean parameters of the environment and,
in particular, the wave-induced flow \citep{Stepanyants89}. In this
case, there is an ambiguous field separation of the physical
variables in the wave field and the medium field. These issues are
rather complicated and analysis continues to this day. For example,
in the MHD context, theory for the wave and background stress energy
tensors is developed based on the exact Lagrangian map in
\citet{Webb05}. We do not go into this debate, but, only aim to get
a feeling of the dynamics of quadratic forms of physical variables.
Therefore, we introduced the concept of energy in solving the
linearized problems as it is usually accepted in hydrodynamic and
magnetodynamic flow studies \citep{Stepanyants89}.

Maximal transient growth (and overreflection) of the wave energy
occurs in the 2D limit (at $k_z=0$). Consequently, the dynamics of
the MHD flow should be defined mainly by SFHs with $k_z/k_x \ll 1$.
At the same time, the transient growth of the  both, P- and
S-Alfv${\rm \acute{e}}$n wave modes increases with decreasing
$\Omega_{A}$. If initially only unidirectional P-AW harmonic is
imposed in the flow, the imbalance degree decreases with the
decrease of $\Omega_{A}$ and the waves become already balanced at
$\Omega_{A}=0.1$. If initially counter-propagating P-AW harmonics
with equal amplitudes are imposed, they maintain the balance
irrespective of the initial phase. Because of the condition
$\Omega_{A} \sim k_x$, SFHs having smaller $k_x$ exhibit stronger
growth. Finally, one can conclude, that the dynamical processes
should be defined by waves having small $k_x$ and $k_z$, i.e. long
streamwise and spanwise wavelengthes, $\lambda_x, \lambda_z \gg
V_A/S$.

\subsection{Impact of the transient growth on the character of nonlinear cascade processes}

The described linear dynamics is expected to have significant
nonlinear consequences. The point is that the linear transient
processes are mainly defined by coefficients $\chi_p$ of equations
(2.13) and (2.14). The coefficients introduce dependency of the
transient growth on the ratio $k_xk_y(\tau)/k^2(\tau)$.
Consequently, the transient growth depends not on the value but on
the orientation of wavevector and takes place when $|\chi_p(\tau)|
\geq \Omega_A$, i.e. when $|k_y(\tau)/k_x| \leq \Omega_A^{-1}$. As
it is outlined at the end of the previous section, this happens in
the vicinity of $\tau \simeq \tau^*$ during the time interval $|\tau
- \tau^*| \leq \Omega_A^{-1}$ (see figures
\ref{fig:fig.4},\ref{fig:fig.6},\ref{fig:fig.7},\ref{fig:fig.11}-\ref{fig:fig.13}).
It is obvious, that the linear transient processes are highly
anisotropic in wavenumber plane. A similar anisotropy exists in
hydrodynamic shear flows  \citet[see for details][]{Horton10}. The
anisotropy changes classical view on the nonlinear cascade
processes: classically, the net action of nonlinear (turbulent)
processes is interpreted as either a direct or inverse cascade.
However, as it is shown in \citet{Horton10}, in hydrodynamic
nonuniform/shear flows, the dominant process is a nonlinear
redistribution over wavevector angles of perturbation spatial
Fourier harmonics. This anisotropic transfers of spectral energy in
the wavenumber space has been coined as \emph{nonlinear transverse
redistribution} (NTR), one can also say -- ``nonlinear transverse
cascade''.

NTR redistributes perturbation harmonics over different quadrants of
the wavenumber plane (e.g., from quadrants where $k_xk_y>0$ to
quadrants where $k_xk_y<0$ or vice versa) and the interplay of this
nonlinear redistribution with linear phenomena (transient growth)
becomes intricate: it can realize either positive or negative
feedback. In the case of positive feedback, the nonlinearity
repopulates transiently growing perturbations and contributes to the
self-sustenance of perturbations. Consequently, NTR naturally
appears as a possible cornerstone of the bypass scenario of
turbulence.

The similarity of the anisotropy of the transient growth in the
hydrodynamic and our MHD cases hints at the similarity of nonlinear
processes. In other words, the transverse cascade should also be
inherent to MHD shear flows. Therefore, the conventional
characterization of MHD turbulence in terms of direct and inverse
cascades, which ignores the transverse cascade, can be misleading
for MHD shear flow turbulence. In principal, the nonlinear
transverse cascade can repopulate the transiently growing wave SFHs
in a MHD shear flow and can acquire a vital role of ensuring the
self-sustenance of the waves. To verify these, consistent with the
bypass concept, processes in MHD shear flows, we already simulated
nonlinear dynamics of the considered here flow system in 2D limit
\citep{Chagelishvili13}. The performed direct numerical simulations
show the existence of subcritical transition to turbulence even in
the 2D case, i.e., show the vitality of the \emph{bypass transition
to turbulence} for the simplest (spectrally stable) plane MHD flow.
The minimal/critical Reynolds number of the subcritical transition
turned out to be about $5000$, i.e. larger than for the hydrodynamic
Couette flow, where $Re_{cr} \sim 350$. At the same time, in the
hydrodynamic case, the transition occurs just in the 3D case, while
in the simplest MHD flow, the subcritical transition and
self-sustenance of turbulence occurs in the 2D case too.

Discussing the character of nonlinear cascade processes, finally,
one has to present the generalising view.

MHD turbulence phenomenon is ubiquitous in nature and is very
important in engineering/industrial applications. So, it is natural
that there is an enormous amount of research devoted to it, starting
with seminal papers \citet{Iroshnikov63} and \citet{Kraichnan65} and
their extensions \citet{Goldreich_Sridhar95,Boldyrev05}. To date,
the main trends, including cases of forced, freely decaying and with
background magnetic field MHD turbulence, established over decades
are thoroughly analyzed in a number of review articles and books
\citep[see e.g.,][]{Biskamp03, Mininni11, Brandenburg_Lazarian13}.
Most of such an analysis commonly focuses on turbulence dynamics in
wavenumber/Fourier space. However, the case of MHD turbulence in
smooth shear flows involves fundamental novelties: the
energy-supplying process for turbulence is the flow nonnormality
induced linear transient growth. The latter \emph{anisotropically
injects energy into turbulence over a broad range of lengthscales},
consequently, rules out the inertial range of the activity of
nonlinearity and leads to complex interplay of linear and nonlinear
processes \citep{Chagelishvili13}. These circumstances give rise to
new type of processes in the turbulence dynamics that are not
accounted for in the main trends of MHD turbulence research.

The essence of the view is the following: anisotropic linear
processes lead to anisotropy of nonlinear processes. Specifically,
the nonlinear transverse cascade (that is anisotropic by definition)
is the result of only \emph{anisotropic} linear coupling/reflection.
For instance, there is a number of papers addressing linear wave
reflection that is caused by parallel gradients of density/magnetic
field \citep[e.g.,][]{Velli89,Matthaeus99,Dmitruk02,
Verdini12,Perez13}. However, this kind of reflection, due to the
absence of the above noted anisotropy in wavenumber space, should
not lead to the nonlinear transverse cascade. A flow configuration,
similar to our magnetized shear flow, is considered by
\citet{Hollweg13}, addressing the wave reflection. The difference is
in the value of beta parameter -- we consider incompressible waves
in high-beta plasma, while, that paper considers compressible waves
in low-beta plasma. The participants of the linear dynamics are
different, however, in the both cases, transient liner
processes/coupling/growth are anisotropic. Consequently, the
nonlinear transverse cascade should also be important for the range
of parameters considered in \citet{Hollweg13}.

\subsection{On application of the proposed approach of the overreflection
to more complex shear flow systems}

Finally, we would like to stress that the present scenario and
mathematical formalism of the overreflection phenomenon is easily
applicable to more complex shear flow systems, including widely
discussed cases of overreflection of spiral-density waves in
astrophysical discs and of internal-gravity waves in stably
stratified atmospheres. The proposed approach describes each
(counter-propagating) wave component by its generalized eigen
variable. In the considered here MHD flow, fortunately, one can use
Els$\ddot{a}$sser variables (see equation 2.12) or its renormalized
version (see equation 2.23) as the generalized eigen variables. This
fact has actually simplified our analysis. As for more complex shear
flow systems, there appears to be some difficulty in finding
generalized eigen variable and construction of fist order
differential equations for each counter-propagating wave. The
difficulty is due to the fact that in complex flow systems ``nominal
frequency'' may depend on the varying/shearwise wavenumber and,
consequently, on time (e.g., as the ``nominal frequency'' of
internal-gravity waves), while in our case considered here Alfv${\rm
\acute{e}}$n waves, the ``nominal frequency'' (Alfv${\rm
\acute{e}}$n frequency) does not depend on the shearwise wavenumber
and is constant. However, this is not a fundamental difficulty -- it
requires just a bit of complicated calculations and gives somewhat
bulky coefficients in dynamical equations. So, to apply the proposed
mathematical approach, one has to find eigen variables of the waves
in the shearless limit, then generalize these eigen variables for
the non-zero shear case and write dynamic equations for them. This
procedure gives the corresponding set of easily foreseeable,
\emph{coupled} first order ordinary differential equations for each
counter-propagating wave.

\section{acknowledgment}

The authors are grateful to Dr. George Mamatsashvili for valuable
help in the preparing of the final version of the manuscript. This
work was supported in part by GNSF grant 31/14.

\bibliographystyle{jpp}

\bibliography{references}

\end{document}